\documentclass{PoS}
\usepackage{bm}

\title{High-energy pulsar light curves in an offset polar cap $B$-field geometry}

\ShortTitle{High-energy pulsar light curves in an offset-PC $B$-field geometry}

\author{\speaker{M. Barnard},$^{a}$ C. Venter$^{a}$
and A. K. Harding$^{b}$\\
\llap{$ˆa$} Centre for Space Research, North-West University (Potchefstroom Campus) \\
Private Bag X6001, Potchefstroom 2520, South Africa\\
\llap{$ˆb$} Astrophysics Science Division, NASA Goddard Space Flight Center,\\
Greenbelt, MD 20771, USA\\
E-mail: \email{monicabarnard77@gmail.com}}

\abstract{The light curves and spectral properties of more than 200 $\gamma$-ray pulsars have been measured in unsurpassed detail in the eight years since the launch of the hugely successful {\it Fermi} Large Area Telescope (LAT) $\gamma$-ray mission. We performed geometric pulsar light curve modelling using static, retarded vacuum, and offset polar cap (PC) dipole $B$-fields (the latter is characterized by a parameter $\epsilon$), in conjunction with standard two-pole caustic (TPC) and outer gap (OG) emission geometries. In addition to constant-emissivity geometric models, we also considered a slot gap (SG) $E$-field associated with the offset-PC dipole $B$-field and found that its inclusion leads to qualitatively different light curves. We therefore find that the assumed $B$-field and especially the $E$-field structure, as well as the emission geometry (magnetic inclination and observer angles), have a great impact on the pulsar's visibility and its high-energy pulse shape. We compared our model light curves to the superior-quality $\gamma$-ray light curve of the Vela pulsar (for energies $>100$~MeV). Our overall optimal light curve fit (with the lowest $\chi^2$ value) is for the retarded vacuum dipole field and OG model. We found that smaller values of $\epsilon$ are favoured for the offset-PC dipole field when assuming \emph{constant} emissivity, and larger $\epsilon$ values are favoured for \emph{variable} emissivity, but not significantly so. When we increased the relatively low SG $E$-fields we found improved light curve fits, with the inferred pulsar geometry being closer to best fits from independent studies in this case. In particular, we found that such a larger SG $E$-field (leading to \emph{variable} emissivity) gives a second overall best fit. This and other indications point to the fact that the actual $E$-field may be larger than predicted by the SG model.}

\FullConference{4th Annual Conference on High Energy Astrophysics in Southern Africa\\
		25-27 August, 2016\\
		Cape Town, South Africa}

\begin{document}



\section{Introduction}\label{sec:intro}

The field of $\gamma$-ray pulsars has been revolutionised by the launch of the \emph{Fermi} Large Area Telescope (LAT; \cite{Atwood2009}). Over the past eight years, \textit{Fermi} has detected over 200 $\gamma$-ray pulsars and has furthermore measured their light curves and spectral characteristics in unprecedented detail. \textit{Fermi}'s Second Pulsar Catalog (2PC; \cite{Abdo2013}) describes the properties of some 117 of these pulsars in the energy range 100~MeV$-$100~GeV. In this paper, we will focus on the GeV band light curves of the Vela pulsar \cite{Abdo2009}, the brightest persistent source in the $\gamma$-ray sky.
 
Physical emission models such as the slot gap (SG; \cite{Muslimov2003}) and outer gap (OG; \cite{Cheng1986,Romani1995}) fall short of fully explaining (global) magnetospheric characteristics, e.g., the particle acceleration and pair production, current closure, and radiation of a complex multi-wavelength spectrum. More recent developments include global magnetospheric models such as the force-free (FF) inside and dissipative outside (FIDO) model \cite{Kalapotharakos2009,Kalapotharakos2014}, the wind models of, e.g., \cite{Petri2011}, and particle-in-cell simulations (PIC; \cite{Cerutti2016a,Cerutti2016b}). Although much progress has been made using these \emph{physical} (or emission) models, \emph{geometric} light curve modeling \cite{Dyks2004,Venter2009,Watters2009,Johnson2014,Pierbattista2015} still presents a crucial avenue for probing the pulsar magnetosphere in the context of traditional pulsar models. The most commonly used emission geometries include the two-pole caustic (TPC; the SG model may be its physical representation; \cite{Dyks2003}) and OG models and may be used to constrain the pulsar geometry (i.e., magnetic inclination angle $\alpha$ and the observer viewing angle $\zeta$ with respect to the spin axis $\boldsymbol\Omega$), as well as the $\gamma$-ray emission region's location and extent. This may provide vital insight into the boundary conditions and help constrain the accelerator geometry of next-generation full radiation models. 

The assumed $B$-field structure is essential for predicting the light curves seen by the observer using geometric models, since photons are expected to be emitted tangentially to the local $B$-field lines in the corotating pulsar frame \cite{Daugherty1982}. Even a small difference in the magnetospheric structure will therefore have an impact on the light curve predictions. Additionally, we have also incorporated an SG $E$-field associated with the offset-PC dipole $B$-field (making this latter case an \textit{emission} model), which allows us to calculate the emissivity $\epsilon_{\nu}$ in the acceleration region in the corotating frame from first principles. 

In this paper, we investigate the impact of different magnetospheric structures (i.e., the static dipole \cite{Griffiths1995}, retarded vacuum dipole (RVD; \cite{Deutsch1955}), and an offset-PC dipole $B$-field solution \cite{Harding2011a,Harding2011b}), as well as the SG $E_\parallel$-field on the pulsar visibility and $\gamma$-ray pulse shape. In combination with the different $B$-field solutions mentioned above, we assume standard TPC and OG emission geometries. In Section~\ref{sec:OPCBfield} we briefly describe the  offset-PC dipole $B$-field and its corresponding SG $E$-field implemented in our code \cite{Dyks2004,Barnard2016}. We also investigate the effect of increasing the $E$-field by a factor of a 100. In Section~\ref{sec:results}, we present our phase plots and model light curves for the Vela pulsar, and we compare our results to previous multi-wavelength studies. Our conclusions follow in Section~\ref{sec:conclusions}.

\section{The Offset-PC Magnetosphere}\label{sec:OPCBfield}

\subsection{$B$-field structure}\label{sub:Bfield}

Several $B$-field structures have been studied in pulsar models, including the static dipole, the RVD (a rotating vacuum magnetosphere which can in principle accelerate particles but do not contain any charges or currents), the FF (filled with charges and currents, but unable to accelerate particles since the accelerating $E$-field is screened everywhere; \cite{Contopoulos1999}), and the offset-PC dipole. The offset-PC dipole solution analytically mimics deviations from the static dipole near the stellar surface and is azimuthally asymmetric, with field lines having a smaller curvature radius over half of the PC (in the direction of the PC offset) compared to those of the other half \cite{Harding2011a,Harding2011b}. Such small distortions in the $B$-field structure can be due to retardation and asymmetric currents, thereby shifting the PCs by small amounts in different directions. A more realistic pulsar magnetosphere, i.e., a dissipative solution \cite{Lichnerowicz1967,Kalapotharakos2012,Li2012,Tchekhovskoy2013,Li2014}, would be one that is intermediate between the RVD and the FF fields. 

The symmetric case involves an offset of both PCs, with respect to the magnetic ($\boldsymbol{\mu}$) axis in the same direction and applies to neutron stars with some interior current distortions that produce multipolar components near the stellar surface \cite{Harding2011a,Harding2011b}. We study the effect of this simpler symmetric case on predicted light curves. The general expression for a symmetric offset-PC dipole $B$-field in spherical coordinates $(r^{\prime},\theta^{\prime},\phi^{\prime})$ in the magnetic frame (indicated by the primed coordinates, where $\hat{{\mathbf{z}}}^{\prime}\parallel{\boldsymbol{\mu}}$) is \cite{Harding2011b}
\begin{eqnarray}\label{eq:Symm-offset}
{{\mathbf{B}}}_{\rm OPCs}^{\prime} & \approx & \frac{\mu^\prime}{r^{\prime3}}\biggl[\cos\theta^{\prime}\hat{{\mathbf{r}}}^{\prime}+\frac{1}{2}(1+a)\sin\theta^{\prime}{\hat{\boldsymbol{\theta}}}^{\prime} - \epsilon\sin\theta^{\prime}\cos\theta^{\prime}
\sin(\phi^{\prime}-\phi_{0}^\prime)
\hat{\boldsymbol{\phi}}^{\prime}\biggr],
\end{eqnarray}
where the symbols have the same meaning as before \cite{Harding2011a,Harding2011b}. We choose the offset direction to be in the $x^\prime-z^\prime$ plane. The $B$-field lines are distorted in all directions, with the distortion depending on parameters $\epsilon$ (related to the magnitude of the shift of the PC from the magnetic axis) and $\phi_0^\prime$ (we choose $\phi^\prime_0=0$ in what follows, with the offset being in the $-x^\prime$ direction). If we set $\epsilon=0$ the symmetric case reduces to a symmetric static dipole. 

The difference between our offset-PC field and a dipole field that is offset with respect to the stellar centre can be most clearly seen by performing a multipolar expansion of these respective fields. An offset dipolar field may be expressed (to lowest order) as the sum of a centred dipole and quadropolar terms $1/r^{\prime4}$). Conversely, our offset-PC field may be written as
\begin{eqnarray}\label{eq:Symm-offset}
{{\mathbf{B}}}_{\rm OPCs}^{\prime}(r^\prime,\theta^\prime,\phi^\prime) & \approx & {\bf B}_{\rm dip}^{\prime}(r^\prime,\theta^\prime)+{\it O}\left(\frac{\epsilon}{r^{\prime3}}\right).
\end{eqnarray}

Therefore, we can see that our offset-PC model (Eq.~[\ref{eq:Symm-offset}]) consists of a centred dipole plus terms of order $a/r^{\prime3}$ or $\epsilon/r^{\prime3}$. Since $a\sim0.2$ and $\epsilon\sim0.2$, the latter terms present perturbations (e.g., poloidal and toroidal effects) to the centred dipole. These perturbed components of the distorted magnetic field were derived under the solenoidality condition $\nabla\cdot B=0$ \cite{Harding2011a,Harding2011b}.

\subsection{Incorporating a corresponding SG $E$-field}\label{sub:Efield}

It is important to take the accelerating $E_\parallel$-field ($E$-field parallel to the local $B$-field,) into account when such expressions are available, since this will modulate the emissivity $\epsilon_\nu$ in the gap as opposed to geometric models where we assume \emph{constant} $\epsilon_\nu$ per unit length in the corotating frame. For the SG case we implement the full $E$-field in the rotational frame corrected for general relativistic (GR) effects (e.g., \cite{Muslimov2003,Muslimov2004a}). 

The low-altitude solution is given by (A.K. Harding 2015, private communication)
\begin{eqnarray}\label{eq:E_low}
E_{\parallel,{\rm low}} & \approx & {-3}{\mathcal{E}_0}\nu_{\rm SG}x^a\Big\{\frac{\kappa}{\eta^4}e_{\rm 1A}\cos\alpha+\frac{1}{4}\frac{\theta_{\rm PC}^{1+a}}{\eta}\big[e_{\rm 2A}\cos\phi_{\rm PC} \nonumber \\
 & & +\frac{1}{4}\epsilon\kappa{e_{\rm 3A}}(2\cos\phi_0^\prime-\cos(2\phi_{\rm PC}-\phi_0^\prime))\big]\sin\alpha\Big\}(1-\xi_\ast^2),
\end{eqnarray}
where the symbols in Eq.~(\ref{eq:E_low}) have the same meaning as in previous works \cite{Muslimov1997,Muslimov2003,Muslimov2004a,Breed2013,Breed2014,Barnard2016}. We choose the negative $x$-axis toward $\boldsymbol{\Omega}$ to coincide with $\phi_{\rm PC}=0$, labeling the ``favourably curved" $B$-field lines.

We approximate the high-altitude SG $E$-field by \cite{Muslimov2004a}
\begin{eqnarray}\label{eq:E_high}
E_{\parallel,{\rm high}} & \approx & -\frac{3}{8}\Big(\frac{\Omega{R}}{c}\Big)^3\frac{B_{\rm 0}}{f(1)}\nu_{\rm SG}x^a\Big\{\Big[1+\frac{1}{3}\kappa\Big(5-\frac{8}{\eta^3_{\rm c}}\Big)+2\frac{\eta}{\eta_{\rm LC}}\Big]\cos\alpha \nonumber \\
 & & +\frac{3}{2}\theta_{\rm PC}H(1)\sin\alpha\cos\phi_{\rm PC}\Big\}(1-\xi_\ast^2).
\end{eqnarray}
The critical scaled radius $\eta_{\rm c}=r_{\rm c}/R$ is where the high-altitude and low-altitude $E$-field solutions are matched, with $r_{\rm c}$ the critical radius, $R$ the stellar radius, $\eta_{\rm LC}=R_{\rm LC}/R$, and $R_{\rm LC}$ the light cylinder radius (where the corotation speed equals the speed of light).

To obtain a general $E$-field valid from $R$ to $R_{\rm LC}$ we use (\cite{Muslimov2004a}; Equation~[59]):
\begin{equation}\label{eq:E_total}
E_{\parallel,{\rm SG}}{\simeq}E_{\parallel,{\rm low}}\exp[-(\eta-1)/(\eta_{\rm c}-1)]+E_{\parallel,{\rm high}}.
\end{equation}
We matched the low-altitude and high-altitude $E$-field solutions by solving $\eta_{\rm c}(P,\dot{P},\alpha,\epsilon,\xi,\phi_{\rm PC})$ on each $B$-field line, where $P$ is the period and $\dot{P}$ its time derivative~\cite{Barnard2016}.

\subsection{Increasing the relatively low $E$-field}\label{sub:100Efield}

In the curvature radiation reaction (CRR, where the energy gain rate equals the CR loss rate) limit, we can determine the CR cutoff of the CR photon spectrum as follows \cite{Venter2010}
\begin{equation}\label{ECRcut}
E_{\rm CR}\sim{4}E_{\parallel,\rm 4}^{3/4}\rho_{\rm curv, 8}^{1/2} \quad {\rm GeV},
\end{equation}
with $\rho_{\rm curv, 8}\sim\rho_{\rm curv}/{10^8}$ cm the curvature radius of the $B$-field line and $E_{\parallel,4}\sim E_{\parallel}/{10^4}$ statvolt cm$^{-1}$. Since the SG $E$-field (see Section~\ref{sub:Efield}) is low (implying a CR cutoff around a few MeV), the phase plots for emission $>100$~MeV display small caustics (Section~\ref{sub:PPLC}) which result in ``missing structure''. Therefore, we investigate the effect on the light curves of the offset-PC dipole $B$-field and SG model combination when we \textit{increase the $E$-field.} As a test we multiply Eq.~(\ref{eq:E_total}) by a factor 100. Using the above expression the estimated cutoff energy for our increased SG $E$-field is now $E_{\rm CR}\sim4$~GeV, which is in the energy range of \textit{Fermi} ($\sim30$~MeV).

\section{Results}\label{sec:results}

\subsection{Phase plots and light curves}\label{sub:PPLC}
\begin{figure}[t]
\includegraphics[width=0.8\textwidth]{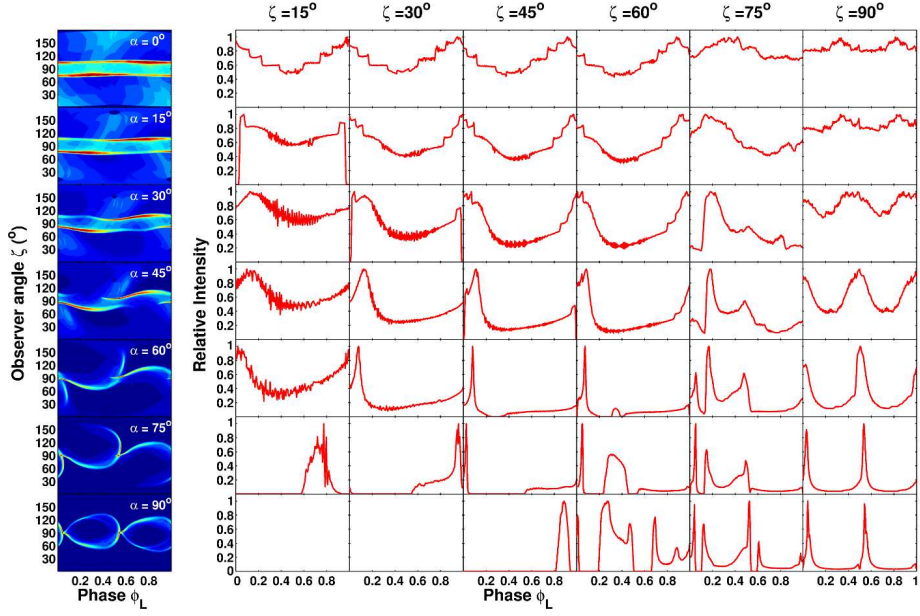}
\centering
\caption{Phase plots (first column) and light curves (second column and onward) for the TPC model assuming an offset-PC dipole field, for a fixed value of $\epsilon=0.18$ and \emph{constant} $\epsilon_{\nu}$. Each phase plot is for a different $\alpha$ value ranging from $0^{\circ}$ to $90^{\circ}$ with a $15^{\circ}$ resolution, and their corresponding light curves are denoted by the solid red lines for different $\zeta$ values, ranging from $15^{\circ}$ to $90^{\circ}$, with a $15^{\circ}$ resolution.}
\label{fig:OffsetEps018}
\end{figure}

As an example we show phase plots and their corresponding light curves for the offset-PC dipole, for both the TPC (assuming uniform $\epsilon_{\nu}$) and SG (assuming \emph{variable} $\epsilon_{\nu}$) models. Figure~\ref{fig:OffsetEps018} is for the TPC model for $\epsilon=0.18$. For larger values of $\alpha$ the caustics extend over a larger range in $\zeta$, with the emission forming a ``closed loop,'' which is also a feature of the static dipole $B$-field at $\alpha=90^{\circ}$. The TPC model is visible at nearly all angle combinations, since some emission occurs below the null charge surface (the geometric surface across which the
charge density changes sign; \cite{GJ1969}) for this model, in contrast to the OG model. However, for $\alpha=90^{\circ}$ and $\zeta$ below $45^{\circ}$ no light curves are visible, i.e., no emission is observed due to the ``closed loop'' structure of the caustics. The TPC light curves exhibit relatively more off-pulse emission than the OG ones. In the TPC model, emission is visible from both magnetic poles, forming double peaks in some cases, whereas in the OG model emission is visible from a single pole. One does obtain double peaks in the OG case, however, when the line of sight crosses the caustic at two different phases.
\begin{figure} 
\includegraphics[width=0.8\textwidth]{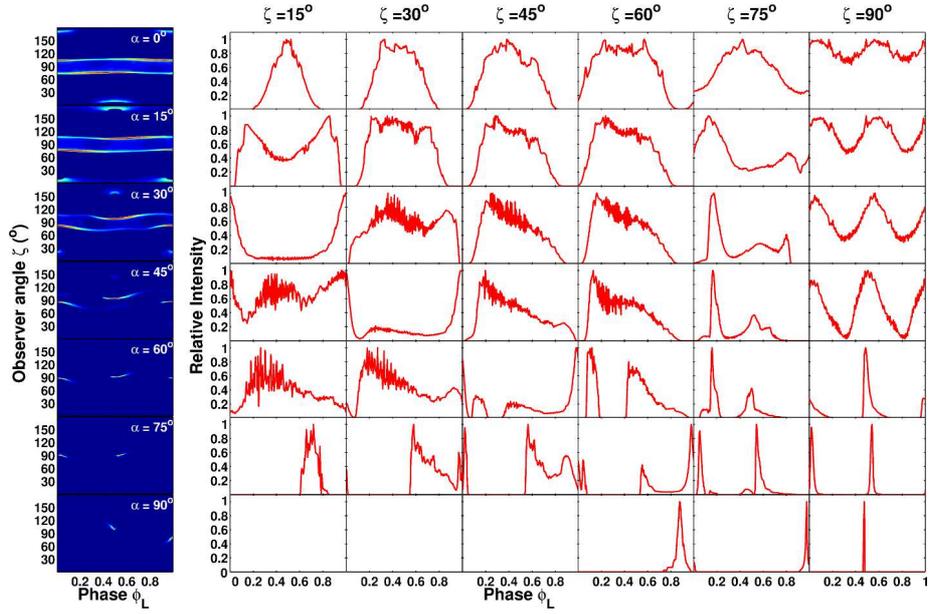}
\centering
\caption{The same as in Figure~\protect\ref{fig:OffsetEps018}, but for the SG model assuming an offset-PC dipole field, for a fixed value of $\epsilon=0.18$ and \emph{variable} $\epsilon_\nu$. The photon energy is above 100~MeV.}
\label{fig:OffsetEps018E}
\end{figure}  
\begin{figure}   
\includegraphics[width=0.8\textwidth]{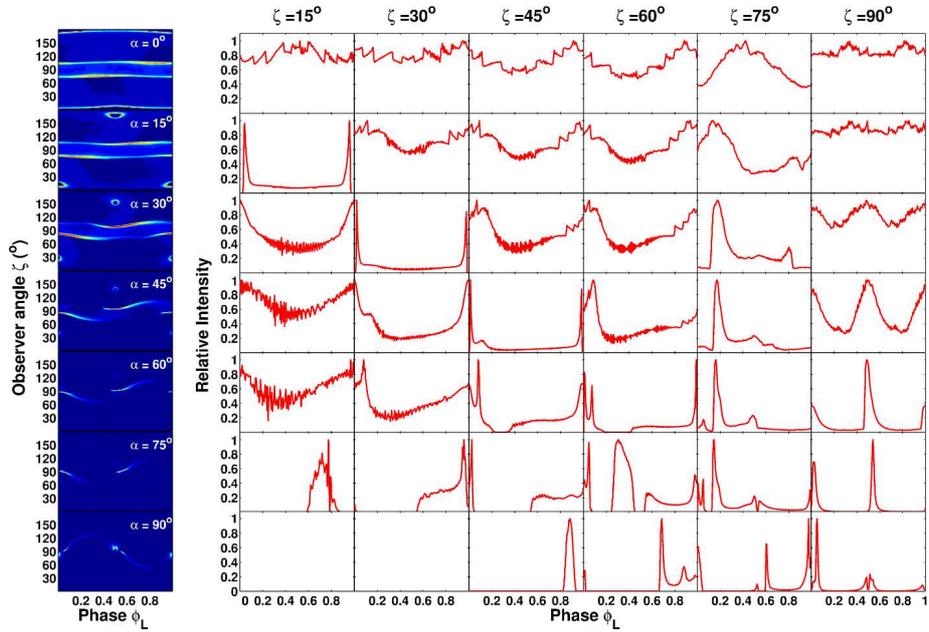}
\centering
\caption{The same as in Figure~\protect\ref{fig:OffsetEps018E}, but for the case where we multiplied $E_\parallel$ by a factor of 100, yielding a CR cutoff of $E_{\rm CR}\sim4$~GeV.}
\label{fig:OffsetEps018GeV}
\end{figure}

If we compare Figure~\ref{fig:OffsetEps018} with the static dipole case (for $\epsilon=0$; not shown), we notice that a larger PC offset $\epsilon$ results in qualitatively different phase plots and light curves, e.g., modulation at small $\alpha$. Also, the caustics occupy a slightly larger region of phase space and seem more pronounced for larger $\epsilon$ and $\alpha$ values. The light curve shapes are also slightly different.

Figure~\ref{fig:OffsetEps018E} is for the offset-PC dipole $B$-field and $\epsilon=0.18$, for a \emph{variable} $\epsilon_\nu$ due to using an SG $E$-field solution (with CR the dominating process for emitting $\gamma$-rays; see Sections~\ref{sub:Efield}). The caustic structure and resulting light curves are qualitatively different for various $\alpha$ compared to the \emph{constant} $\epsilon_\nu$ case. The caustics appear smaller and less pronounced for larger $\alpha$ values (since $E_\parallel$ becomes lower as $\alpha$ increases), and extend over a smaller range in $\zeta$. If we compare Figure~\ref{fig:OffsetEps018E} with the case for $\epsilon=0$ (for \emph{variable} $\epsilon_\nu$; not shown) we note a new emission structure close to the PCs for small values of $\alpha$ and $\zeta\approx(0^\circ,180^\circ)$. This reflects the boosted $E_\parallel$-field on the ``favourably curved'' $B$-field lines. In Figure~\ref{fig:OffsetEps018E} a smaller region in phase space filled. The light curves generally display only one broad peak with less off-peak emission compared to Figure~\ref{fig:OffsetEps018}. As $\alpha$ and $\zeta$ increase, more peaks become visible, with emission still visible from both poles as seen for larger $\alpha$ and $\zeta$ values, e.g., $\alpha=75^{\circ}$ and $\zeta=75^{\circ}$.

If we compare Figures~\ref{fig:OffsetEps018} with ~\ref{fig:OffsetEps018E}, we notice that when we take $E_{\parallel}$ into account, the phase plots and light curves change considerably. For example, for $\alpha=90^{\circ}$ in the \emph{constant} $\epsilon_\nu$ case, a ``closed loop'' emission pattern is visible in the phase plot, which is different compared to the small ``wing-like'' emission pattern in the \emph{variable} $\epsilon_\nu$ case. \textit{Therefore, we see that both the $B$-field and $E$-field have an impact on the predicted light curves.} This small ``wing-like'' caustic pattern is due to the fact that we only included photons in the phase plot with energies $>100$~MeV. Given the relatively low $E$-field, there are only a few photons with energies exceeding $100$~MeV. 

In Figure~\ref{fig:OffsetEps018GeV} we present the phase plots and light curves for the SG $E$-field (increased by a factor 100) for the offset-PC dipole and SG model solution, with $\epsilon=0.18$. If we compare Figure~\ref{fig:OffsetEps018GeV} with Figure~\ref{fig:OffsetEps018E} we notice that more phase space is filled by caustics, especially at larger $\alpha$. At $\alpha=90^\circ$ the visibility is again enhanced. The caustic structure becomes wider and more pronounced, with extra emission features arising as seen at larger $\alpha$ and $\zeta$ values. This leads to small changes in the light curve shapes. At smaller $\alpha$ values, the emission around the PC forms a circular pattern that becomes smaller as $\alpha$ increases. These rings around the PCs become visible since the low $E$-field is boosted, leading to an increase in bridge (region between the first and second peak of a light curve) emission as well as higher signal-to-noise ratio. At low $\alpha$ the background becomes feature-rich, but not at significant intensities, however.


\subsection{Comparison of best-fit parameters for different models}\label{sub:compmod}

We next follow the same approach as a previous study \cite{Pierbattista2015} to compare the various optimal solutions of the different models. We determine the difference between the scaled $\chi^2$ of the optimal model, $\xi_{\rm opt}^{2}$, and the other models ($\xi^{2}$) using
\begin{eqnarray}\label{eq:difference-chi2}
\Delta\xi^{2} & =\xi^{2}-\xi_{{\rm opt}}^{2}=N_{{\rm dof}}\left(\chi^{2}/\chi_{{\rm opt}}^{2}-1\right).\end{eqnarray}
with degrees of freedom $N_{\rm dof}=96$. We considered two approaches: we found the best fit (i) per $B$-field and model combination ($\Delta\xi^{2}_{\rm B}$), and (ii) overall (for all $B$-field and model combinations, $\Delta\xi^{2}_{\rm all}$)\footnote{We therefore first scale the $\chi^2$ values using the optimal value obtained for a particular $B$-field, and second we scale these using the overall optimal value irrespective of $B$-field.}.

\begin{figure}[t]
\includegraphics[width=1.0\textwidth]{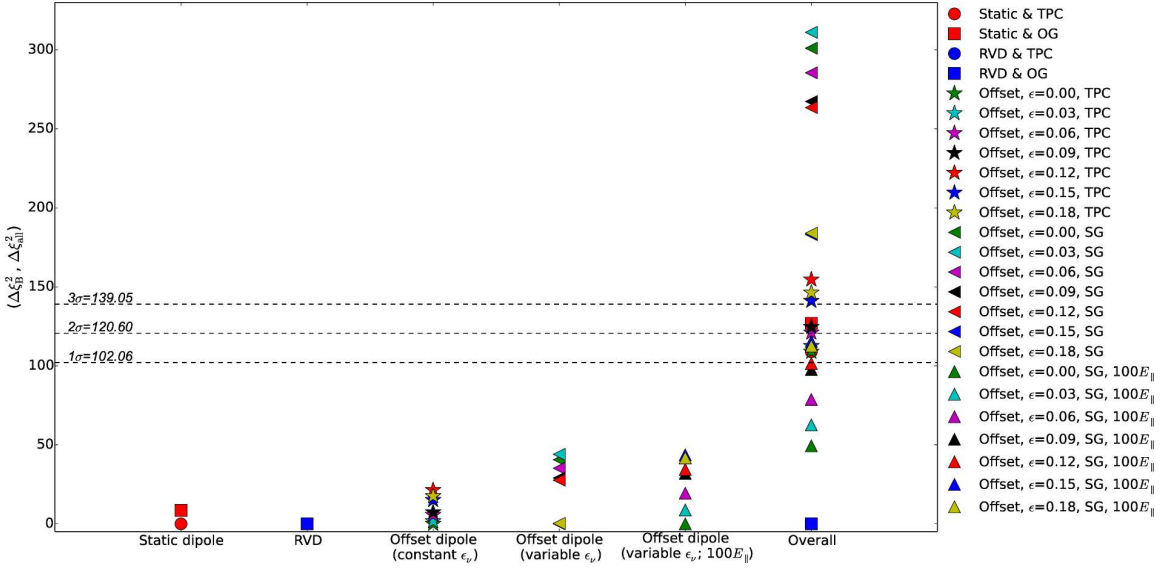}
\caption{Comparison of the relative goodness of the fit of solutions obtained for each $B$-field and geometric model combination, including the case of $100E_\parallel$, as well as all combinations compared to the overall best fit, i.e., RVD $B$-field and OG model (shown on the $x$-axis). The difference between the optimum and alternative model for each $B$-field is expressed as $\Delta\xi^{2}_{\rm B}$, and for the overall fit as $\Delta\xi^{2}_{\rm all}$ (shown on the $y$-axis). The horizontal dashed lines indicate the $1\sigma$, $2\sigma$, and $3\sigma$ confidence levels. Circles and squares refer to the TPC and OG models for both the static dipole and RVD. The stars refer to the TPC (\emph{constant} $\epsilon_{{\rm \nu}}$) and the left pointing triangles present the SG (\emph{variable} $\epsilon_{{\rm \nu}}$) model for the offset-PC dipole field, for the different $\epsilon$ values. The upright triangles refer to our SG model and offset-PC dipole case for a larger $E$-field ($100E_\parallel$). The last column shows our overall fit comparison (see legend for symbols).}
\label{fig:ModelComparisonB}
\end{figure}

In Figure~\ref{fig:ModelComparisonB} we label the different $B$-field structures assumed in the various models as well as the overall comparison along the $x$-axis, and plot $\Delta\xi^{2}_{\rm B}$ and $\Delta\xi^{2}_{\rm all}$ on the $y$-axis. We represent the TPC geometry with a circle, the OG with a square, and for the offset-PC dipole field we represent the various $\epsilon$ values for \emph{constant} $\epsilon_{\nu}$ by different coloured stars, for \emph{variable} $\epsilon_{\nu}$ by different coloured left pointing triangles, and for the case of $100E_\parallel$ by different coloured upright triangles, as indicated in the legend. The dashed horizontal lines indicate the confidence levels we obtained for $N_{\rm dof}=96$ degrees of freedom. These confidence levels are used as indicators of when to reject or accept an alternative fit compared to the optimum fit.

For the static dipole field the TPC model gives the optimum fit and the OG model lies within $1\sigma$ of this fit, implying that the OG geometry may provide an acceptable alternative fit to the data in this case. For the RVD field the TPC model is significantly rejected beyond the $3\sigma$ level (not shown on plot), and the OG model is preferred. We show three cases for the offset-PC dipole field, including the TPC model assuming \emph{constant} $\epsilon_{\nu}$, the SG model assuming \emph{variable} $\epsilon_{\nu}$, and the latter is for an $E_\parallel$-field multiplied by a factor of $100$. The optimal fits for the offset-PC dipole field and TPC model reveal that a smaller offset is generally preferred for \emph{constant} $\epsilon_{\nu}$, while a larger offset is preferred for \emph{variable} $\epsilon_{\nu}$ (but not significantly), with all alternative fits falling within $1\sigma$ of these. However, when we  increase $E_\parallel$-field, a smaller offset is preferred for the SG and \emph{variable} $\epsilon_\nu$ case. When we compare all model and $B$-field combinations with the overall best fit (i.e., rescaling the $\chi^2$ values of all combinations using the optimal fit involving the RVD $B$-field and OG model), we notice that the static dipole and TPC model falls within $2\sigma$, whereas the static OG model lies within $3\sigma$. We also note that the usual offset-PC dipole $B$-field and TPC model combination (for all $\epsilon$ values) is above $1\sigma$ (with some fits $<2\sigma$), but the offset-PC dipole $B$-field and SG model combination (for all $\epsilon$ values) is significantly rejected ($>3\sigma$). However, the case of the offset-PC dipole field and a higher SG $E_\parallel$-field for all $\epsilon$ values leads to a recovery, since all the fits fall within $1\sigma$ or $2\sigma$ and delivers an overall optimal fit for $\epsilon=0$, second only to the RVD and OG model fit. 

\begin{figure}[t]
\centering
\includegraphics[width=.8\textwidth]{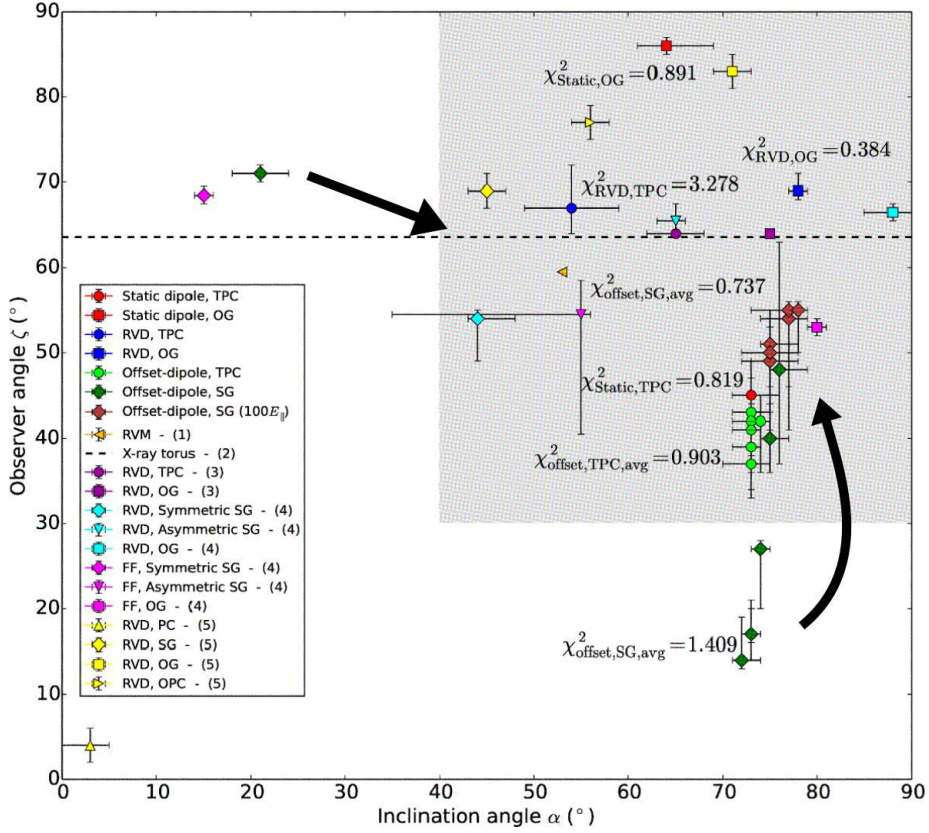}
\caption{Comparison between the best-fit $\alpha$ and $\zeta$, with errors, obtained from this and other studies, e.g., (1) \cite{Johnston2005}, (2) \cite{Ng2008}, (3) \cite{Watters2009}, (4) \cite{DeCesar2013}, and (5) \cite{Pierbattista2015}. The unscaled $\chi^2$~($\times10^5$) value of our fits are also indicated. For the offset-PC dipole, for both the TPC and SG models we indicate the average $\chi^2$ value over the range of $\epsilon$. We also show our fits for the offset-PC dipole and SG model case with the increased $E_\parallel$-field. The two black arrows indicate the shift of the best fits to larger $\alpha$ and $\zeta$ if we increase our SG $E$-field by a factor of 100. The shaded region contains all the fits that cluster at larger $\alpha$ and $\zeta$ values.}
\label{fig:Comparison_alphazeta}
\end{figure}

Several multi-wavelength studies have been performed for Vela, using the radio, X-ray, and $\gamma$-ray data, in order to find constraints on $\alpha$ and $\zeta$. We graphically summarise the best-fit $\alpha$ and $\zeta$, with errors, from this and other works in Figure~\ref{fig:Comparison_alphazeta}. We notice that the best fits generally prefer a large $\alpha$ or $\zeta$ or both. It is encouraging that many of the best-fit solutions lie near the $\zeta$ inferred from the pulsar wind nebula (PWN) torus fitting \cite{Ng2008}, notably for the RVD $B$-field. A significant fraction of fits furthermore lie near the $\alpha-\zeta$ diagonal, i.e., they prefer a small impact angle, most probably due to radio visibility constraints \cite{Johnson2014}. For an isotropic distribution of pulsar viewing angles, one expects $\zeta$ values to be distributed as $\sin(\zeta)$ between $\zeta=[0^\circ,90^\circ]$, i.e., large $\zeta$ values are much more likely than small $\zeta$ values, which seems to agree with the large best-fit $\zeta$ values we obtain. There seems to be a reasonable correspondence between our results obtained for geometric models and those of other authors, but less so for the offset-PC dipole $B$-field, and in particular for the SG $E$-field case. The lone fit near $(20^\circ,70^\circ)$ may be explained by the fact that a very similar fit, but one with slightly worse $\chi^2$, is found at $(50^\circ,80^\circ)$. If we discard the non-optimal TPC / SG fits, we see that the optimal fits will cluster near the other fits at large $\alpha$ and $\zeta$. Although our best fits for the offset-PC dipole $B$-field are clustered, it seems that increasing $\epsilon$ leads to a marginal decrease in $\zeta$ for the TPC model (light green) and opposite for SG (dark green), but not significantly. For our increased SG $E$-field case (brown) we note that the fits now cluster inside the gray area above the fits for the static dipole and TPC, and offset-PC dipole for both the TPC and SG geometries. 

\section{Conclusions}\label{sec:conclusions}

We investigated the impact of different magnetospheric structures (i.e., static dipole, RVD, and a symmetric offset-PC dipole fields) on predicted $\gamma$-ray pulsar light curve characteristics. For the offset-PC dipole field we only considered the TPC (assuming uniform $\epsilon_{\nu}$) and SG (modulating the $\epsilon_{\nu}$ using the $E$-field which is corrected for GR effects up to high altitudes) models. We concluded that the magnetospheric structure and emission geometry have an important effect on the predicted $\gamma$-ray pulsar light curves. \textit{However, the presence of an $E$-field may have an even greater effect than small changes in the $B$-field and emission geometries}.

We fit our model light curves to the observed {\it Fermi}-measured Vela light curve for each $B$-field and geometric model combination. We found that the RVD field and OG model combination fit the observed light curve the best for $(\alpha,\zeta)=({78_{-1}^{+1}}^\circ,{69_{-1}^{+2}}^\circ)$ and an unscaled $\chi^2=3.84\times{10}^4$. As seen in Figure~\ref{fig:ModelComparisonB}, for the RVD field an OG model is significantly preferred over the TPC model, given the characteristically low off-peak emission. For the other field and model combinations there was no significantly preferred model (per $B$-field), since all the alternative models may provide an acceptable alternative fit to the data, within $1\sigma$. The offset-PC dipole field for \emph{constant} $\epsilon_{\nu}$ favoured smaller values of $\epsilon$, and for \emph{variable} $\epsilon_{\nu}$ larger $\epsilon$ values, but not significantly so ($<1\sigma$). When comparing all cases (i.e., all $B$-fields), we noted that the offset-PC dipole field for \emph{variable} $\epsilon_{\nu}$ was significantly rejected ($>3\sigma$). 

Since we wanted to compare our model light curves to \textit{Fermi} data we increased the usual low SG $E$-field by a factor of 100\footnote{This number is not unreasonable, especially in light of the observed spectral high-energy spectral cutoffs. Since pulsars have high local $B$-field strengths and we expect that $E_{\parallel}\leq B$, such high $E$-fields are realistic. A larger gap width is also likely, and this will further increase $E_{\parallel}$.} (leading to a spectral cutoff $E_{\rm CR}\sim4$~GeV). The increased $E$-field also had a great impact on the phase plots, e.g., extended caustic structures and new emission features as well as different light curve shapes emerged. We noted that a smaller $\epsilon$ was again (as in the TPC case) preferred, although not significantly ($<1\sigma$). When we compared this case to the other $B$-field and model combinations, we found statistically better $\chi^2$ fits for all $\epsilon$ values, with an optimal fit at $\alpha={75^{+3}_{-1}}^{\circ}$ and $\zeta={51^{+2}_{-5}}^{\circ}$ for $\epsilon=0$ being second in quality only to the RVD and OG model fit. 

We found reasonable correspondence between our results obtained for geometric models and those of other independent studies. We noted that the optimal fits generally clustered near the other fits at large $\alpha$ and $\zeta$. For our increased SG $E$-field and offset-PC dipole combination, we noted that these fits now clustered at larger $\alpha$ and $\zeta$.

There have been several indications that \textit{the SG $E$-field may be larger than initially thought, as confirmed by this study.} (i) Population synthesis studies found that the SG $\gamma$-ray luminosity may be too low, pointing to an increased $E$-field and / or particle current through the gap, e.g., \cite{Pierbattista2015}. (ii)~If the $E$-field is too low, one is not able to reproduce the observed spectral cutoffs of a few GeV (Section~\ref{sub:100Efield}; \cite{Abdo2013}). (iii) A larger $E$-field (increased by a factor of 100) led to statistically improved $\chi^2$ fits with respect to the light curves. (iv) The inferred best-fit $\alpha$ and $\zeta$ parameters for this $E$-field clustered near the best fits of independent studies. (v) A larger SG $E$-field also increased the particle energy gain rates leading to CRR being reached close to the stellar surface. 

Independent multi-wavelength studies have considered many other pulsars, in addition to the Vela pulsar. For example, Ng \& Romani~\cite{Ng2004,Ng2008} used torus and jet fitting to constrain $\zeta$ of a few X-ray pulsars, and obtained a consistent value of $\zeta=63.6^{+0.07}_{-0.05}$. Johnson et al.~\cite{Johnson2014} and Pierbattista et al.~\cite{Pierbattista2015} fitted the radio and $\gamma$-ray light curves of millisecond and younger pulsar populations respectively using standard geometric models. DeCesar et al.~\cite{DeCesar2013} constrained the $\alpha$ and $\zeta$ angles of a handful of pulsars using standard emission geometries coupled with the FF $B$-field. Overall, there seems to be reasonable consistency between the best-fit geometries derived using the various models.

A number of studies have lastly considered signatures in the polarisation domain for different $B$-field geometries, radiation mechanisms, and emission sites, e.g., \cite{Dyks2004,Cerutti2016b,Harding2017}. This avenue may well prove very important in future to aid in differentiating between the various pulsar models, in addition to spectral and light curve measurements.

\acknowledgments
We thank Marco Pierbattista, Tyrel Johnson, Lucas Guillemot, and Bertie Seyffert for fruitful discussions. This work is based on the research supported wholly / in part by the National Research Foundation of South Africa (NRF; Grant Numbers 87613, 90822, 92860, 93278, and 99072). The Grantholder acknowledges that opinions, findings and conclusions or recommendations expressed in any publication generated by the NRF supported research is that of the author(s), and that the NRF accepts no liability whatsoever in this regard. A.K.H.\ acknowledges the support from the NASA Astrophysics Theory Program. C.V.\ and A.K.H.\ acknowledge support from the \textit{Fermi} Guest Investigator Program.


\begin{thebibliography}{99}

\bibitem{Abdo2009} A.~A. Abdo, M. Ackermann, W.~B. Atwood \emph{et al.}, \emph{Fermi Large Area Telescope Observations of the Vela Pulsar}, \emph{ApJ} {\bf 696} 1084 (2009).

\bibitem{Abdo2013} A.~A. Abdo, M. Ajello, A. Allafort \emph{et al.}, \emph{The Second Fermi Large Area Telescope Catalog of Gamma-Ray Pulsars}, \emph{ApJS} {\bf 208} 17 (2013).

\bibitem{Atwood2009} W.~B. Atwood, A.~A. Abdo, M. Ackermann \emph{et al.}, \emph{The Large Area Telescope on the Fermi Gamma-Ray Space Telescope Mission}, \emph{ApJ} {\bf 697} 1071 (2009).

\bibitem{Barnard2016} M. Barnard, C. Venter, and A.~K. Harding, \emph{The Effect of an Offset Polar Cap Dipolar Magnetic Field on the
Modeling of the Vela Pulsar's Gamma-Ray Light Curves}, \emph{ApJ} {\bf 832} 107 (2016).

\bibitem{Breed2013} M. Breed, C. Venter, A.~K. Harding, and T.~J. Johnson, \emph{Implementation of an Offset-dipole Magnetic Field in a Pulsar Modelling Code}, in proceedings of \emph{SAIP2013: the 58$^{th}$ Ann. Conf. of the SA Institute of Physics} ed. R. Botha and T. Jili, 350 (2014).

\bibitem{Breed2012} M. Breed, C. Venter, A.~K. Harding, and T.~J. Johnson, \emph{The Effect of Different Magnetospheric Structures on Predictions of Gamma-Ray Pulsar Light Curves}, in proceeding of \emph{SAIP2012: the 57$^{th}$ Ann. Conf. of the SA Institute of Physics} ed. J. Janse van Rensburg, 316 (2015).

\bibitem{Breed2014} M. Breed, C. Venter, A.~K. Harding, and T.~J. Johnson, \emph{The Effect of an Offset-dipole Magnetic Field on the Vela Pulsar's Gamma-Ray Light Curves}, in proceedings of \emph{SAIP2014: the 59$^{th}$ Ann. Conf. of the SA Institute of Physics} ed. C. Engelbrecht and S. Karataglidis, 311 (2015).

\bibitem{Cerutti2016a} B. Cerutti, A.~A. Philippov, and A. Spitkovsky, \emph{Modelling High-Energy Pulsar Light Curves from First Principles}, \emph{MNRAS} {\bf 457} 2401 (2016).

\bibitem{Cerutti2016b} B. Cerutti, J. Mortier, and A.~A. Philippov, \emph{Polarized Synchrotron Emission from the Equatorial Current Sheet in Gamma-Ray Pulsars}, \emph{MNRAS} {\bf 463} L89 (2016).

\bibitem{Cheng1986} K.~S. Cheng, C. Ho, and M. Ruderman, \emph{Energetic Radiation from Rapidly Spinning Pulsars. I $-$ Outer Magnetosphere Gaps. II $-$ VELA and Crab}, \emph{ApJ} {\bf 300} 500 (1986).

\bibitem{Contopoulos1999} I. Contopoulos, D. Kazanas, and C. Fendt, \emph{The Axisymmetric Pulsar Magnetosphere}, \emph{ApJ} {\bf 511} 351 (1999).

\bibitem{Daugherty1982} J.~K. Daugherty, and A.~K. Harding, \emph{Electromagnetic Cascades in Pulsars}, \emph{ApJ} {\bf 252} 337 (1982).

\bibitem{DeCesar2013} M.~E. DeCesar, \emph{Using Fermi Large Area Telescope Observations to Constrain the Emission and Field Geometries of Young Gamma-Ray Pulsars and to Guide Millisecond Pulsar Searches}, \emph{PhD thesis}, Univ. of Maryland, College Park (2013).

\bibitem{Deutsch1955} A.~J. Deutsch, \emph{The Electromagnetic Field of an Idealized Star in Rigid Rotation in Vacuo}, \emph{AnAp} {\bf 18} 1 (1955).

\bibitem{Dyks2003} J. Dyks, and B. Rudak, \emph{Two-Pole Caustic Model for High-Energy Light Curves of Pulsars}, \emph{ApJ} {\bf 598} 1201 (2003).

\bibitem{Dyks2004} J. Dyks, A.~K. Harding, and B. Rudak, \emph{Relativistic Effects and Polarization in Three High-Energy Pulsar Models}, \emph{ApJ} {\bf 606} 1125 (2004).

\bibitem{GJ1969} P. Goldreich, and W. H. Julian, \emph{Pulsar Electrodynamics}, \emph{ApJ} {\bf 157} 869 (1969).

\bibitem{Griffiths1995} D.~J. Griffiths, \emph{Introduction to Electrodynamics}, $3^{\rm rd}$ ed.; San Francisco: Pearson Benjamin Cummings (1995).

\bibitem{Harding2011a} A.~K. Harding, and A.~G. Muslimov, \emph{Pulsar Pair Cascades in a Distorted Magnetic Dipole Field}, \emph{ApJL} {\bf 726} L10 (2011).

\bibitem{Harding2011b} A.~K. Harding, and A.~G. Muslimov, \emph{Pulsar Pair Cascades in Magnetic Fields with Offset Polar Caps}, \emph{ApJ} {\bf 743} 181 (2011).

\bibitem{Harding2017} A.~K. Harding, and C. Kalapotharakos, \emph{in preparation}, (2017).

\bibitem{Johnson2014} T.~J. Johnson, C. Venter, A.~K. Harding \emph{et al.}, \emph{Constraints on the Emission Geometries and Spin Evolution of Gamma-Ray Millisecond Pulsars}, \emph{ApJS} {\bf 213} 6 (2014).

\bibitem{Johnston2005} S. Johnston, G. Hobbs, S. Vigeland \emph{et al.}, \emph{Evidence For Alignment of the Rotation and Velocity Vectors in Pulsars}, \emph{MNRAS} {\bf 364} 1397 (2005).

\bibitem{Kalapotharakos2009} C. Kalapotharakos, and I. Contopoulos, \emph{Three-dimensional Numerical Simulations of the Pulsar Magnetosphere: Preliminary Results}, \emph{A\&{A}} {\bf 496} 495 (2009).

\bibitem{Kalapotharakos2014} C. Kalapotharakos, A.~K. Harding, and D. Kazanas, \emph{Gamma-Ray Emission in Dissipative Pulsar Magnetospheres: From Theory to Fermi Observations}, \emph{ApJ} {\bf 793} 97 (2014).

\bibitem{Kalapotharakos2012} C. Kalapotharakos, D. Kazanas, A.~K. Harding, and I. Contopoulos, \emph{Toward a Realistic Pulsar Magnetosphere}, \emph{ApJ} {\bf 749} 2 (2012).

\bibitem{Li2014} J.~G. Li, \emph{Electromagnetic and Radiative Properties of Neutron Star Magnetospheres}, PhD thesis, Princeton Univ., New Jersey (2014).

\bibitem{Li2012} J. Li, A. Spitkovsky, and A. Tchekhovskoy, \emph{Resistive Solutions for Pulsar Magnetospheres}, \emph{ApJ} {\bf 746} 60 (2012).

\bibitem{Lichnerowicz1967} A. Lichnerowicz, \emph{Relativistic Hydrodynamics and Magnetohydrodynamics}, $1^{\rm st}$ ed.; New York: Benjamin, Inc. (1967).

\bibitem{Lowrie2011} W. Lowrie, \emph{A Student's Guide to Geophysical Equations}, $1^{\rm st}$ ed.; Cambridge University Press (2011).

\bibitem{Muslimov1997} A.~G. Muslimov, and A.~K. Harding, \emph{Toward the Quasi-Steady State Electrodynamics of a Neutron Star}, \emph{ApJ} {\bf 485} 735 (1997).

\bibitem{Muslimov2003} A.~G. Muslimov, and A.~K. Harding, \emph{Extended Acceleration in Slot Gaps and Pulsar High-Energy Emission}, \emph{ApJ} {\bf 588} 430 (2003).

\bibitem{Muslimov2004a} A.~G. Muslimov, and A.~K. Harding, \emph{High-Altitude Particle Acceleration and Radiation in Pulsar Slot Gaps}, \emph{ApJ} {\bf 606} 1143 (2004).

\bibitem{Ng2004} C.-Y. Ng, and R.~W. Romani, \emph{Fitting Pulsar Wind Tori.}, \emph{ApJ} {\bf 601} 479 (2004).

\bibitem{Ng2008} C.-Y. Ng, and R.~W. Romani, \emph{Fitting Pulsar Wind Tori.\ II.\ Error Analysis and Applications}, \emph{ApJ} {\bf 673} 411 (2008).

\bibitem{Petri2011} J. P{\'e}tri, and G. Dubus, \emph{Implication of the Striped Pulsar Wind Model for Gamma-Ray Binaries}, \emph{MNRAS} {\bf 417} 532 (2011).

\bibitem{Pierbattista2015} M. Pierbattista, A.~K. Harding, I.~A. Grenier \emph{et al.}, \emph{Light-curve Modelling Constraints on the Obliquities and Aspect Angles of the Young Fermi Pulsars}, \emph{A\&{A}} {\bf 575} A3 (2015).

\bibitem{Romani1995} R.~W. Romani, and I.-A. Yadigaroglu, \emph{Gamma-Ray Pulsars: Emission Zones and Viewing Geometries}, \emph{ApJ} {\bf 438} 314 (1995).

\bibitem{Tchekhovskoy2013} A. Tchekhovskoy, A. Spitkovsky, and J.~G. Li, \emph{Time-dependent 3D Magnetohydrodynamic Pulsar Magnetospheres: Oblique Rotators}, \emph{MNRAS} {\bf 435} L1 (2013).

\bibitem{Venter2010} C. Venter, and O.~C. de Jager, \emph{Accelerating High-energy Pulsar Radiation Codes}, \emph{ApJ} {\bf 725} 1903 (2010).

\bibitem{Venter2009} C. Venter, A.~K. Harding, and L. Guillemot, \emph{Probing Millisecond Pulsar Emission Geometry Using Light Curves from the Fermi/Large Area Telescope}, \emph{ApJ} {\bf 707} 800 (2009).

\bibitem{Watters2009} K.~P. Watters, R.~W. Romani, P. Weltevrede, and S. Johnston, \emph{An Atlas for Interpreting Gamma-Ray Pulsar Light Curves}, \emph{ApJ} {\bf 695} 1289 (2009).

\end{thebibliography}
\end{document}